\documentclass{iau}
\usepackage{graphicx}

\title[IAUS291.~~Polarization sounding of pulsar magnetosphere] 
{Polarization sounding of the pulsar magnetosphere}

\author[O. M. Ulyanov, A. I. Shevtsova \& A. A. Seredkina]   
{O. M. Ulyanov,
A. I. Shevtsova \and
A. A. Seredkina}

\affiliation{Department of Astrophysics, Institute of Radio Astronomy of NAS of Ukraine, \\ Krasnoznamennaya str.4,
Kharkov 61002, Ukraine\\ email: {\tt oulyanov@rian.kharkov.ua, alice.shevtsova@gmail.com, seredkina.a@gmail.com} } 

\pubyear{2012}
\volume{291}  
\jname{\mbox{Neutron Stars and Pulsars: Challenges and
Opportunities after 80 years}}
\editors{J. van Leeuwen, ed.}
\begin{document}

\maketitle

\begin{abstract}
The possibility of a polarization sounding of the pulsar
magnetosphere is examined, using intrinsic pulsar emission as a probe signal,
for modern radio telescopes operating in the meter and decameter
wavelength range. Different models of the pulsar magnetosphere at
altitudes higher than a radius of critical polarization are used.
The propagation medium besides magnetosphere is described by the
stratified model, in which each layer has its own density of free
electrons and vector of magnetic induction, as well as the spatial and
temporal fluctuation scales of these parameters.

The frequency dependence of the polarization parameters of the
pulsar radio emission, obtained in the broad band for a selected
pulse phase, will enable a sounding deep into the pulsar
magnetosphere.  


\end{abstract}

\firstsection 
\section{Introduction}

The pulsar magnetosphere is a region where radio emission is
generated. Studying the properties of radio-emitting region is of
primary importance for understanding both the nature of pulsar
radio emission, and the nature of the neutron star as a whole.
There are several models of pulsar magnetosphere, but still there
is no unified explanation of physical processes occurring in a
pulsar magnetosphere.

The main purpose of this work is to sound the pulsar magnetosphere
by studying the polarization properties and characteristics of a
pulsar radio emission. The decameter range is the most difficult for
polarization studies: there the influence of all known
propagation effects is becomes apparent with the highest contrast.
It is in this range the anomalously powerful pulses from
pulsars B0809+74, B0943+10, B0950+08, B1133+16 
(\cite[Ul'yanov \etal\ 2006]{Ul'yanov_etal06}) 
and the giant pulses from PSR B0531+21 
(\cite[Popov \etal\ 2006]{Popov_etal06}) 
are registered.

Furthermore, due to the Faraday effect the bandwidth of intensity
modulation of the elliptically polarized radiation in this range
narrows down to several tens of kilohertz. It is supposed that this
property of radio emission of pulsars (REP) will be used to obtain
the polarization parameters of REP via observations on radio
telescope UTR-2 which is composed from linearly polarized
broadband dipoles. For a moment the observations of REP
polarization using this effect were reported for the
radio-telescopes DKR 1000 and BCA 100
 (\cite[Suleymanova \& Pugachev 2002]{SuleymanovaPugachev02}). 
Determination of the polarization parameters requires consideration of
both direct and inverse problems. To achieve the assigned task we
consider a model of polarized radiation and a model of the propagation
medium. Solving the inverse problem allows for the recovery of Stokes
parameters in the reference frame of the pulsar using registered
REP at the receiver side. In our work the direct and inverse
problems are solved as applied to the case of UTR-2 telescope.

\begin{figure}[b]
\begin{center}
 \includegraphics[width=5.3in]{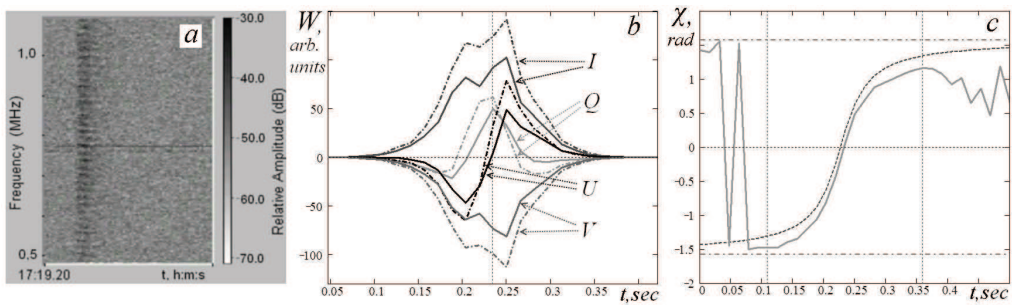}
 \caption{ a)  the dynamic spectrum of the elliptically polarized radiation from PSR B0809+74, detected using UTR-2
 ($F_c$  = 23.7 MHz, $\Delta F $ = 1.53 MHz); b)  the Stokes parameters $I, Q, U, V;$  c) the position angle $\chi$.}
   \label{fig1}
\end{center}
\end{figure}

\section{Model}

To solve the assigned task we model a medium of propagation as a
layered structure. The most important parts of the propagation
medium are the upper pulsar magnetosphere, the interstellar medium
(ISM), the interplanetary medium, the Earth's ionosphere and the
underlying surface of a radio telescope. The layers are
characterized by their own transmission coefficients and by their
own the longitudinal and transverse magnetic field components on
the line of sight, etc. We assume that propagation of a radiation
along a line of sight in each layer is described by a eikonal
equation, which includes separate refractive coefficients for the
ordinary and extraordinary waves. These coefficients we determine
via the mean values of the electron concentration in the selected
layer on the line of sight and the average values of magnetic
field vector component parallel to the line of sight.

At present stage the propagation of the radio waves in the ISM is
considered, since modeling of this medium is the most simple. A
radiation point source located at infinitely large distance from
the receiving antennas and emitting elliptically polarized
radiation in a wide frequency range is considered. We will assume
that at distances of a critical radius of polarization the two
orthogonal modes of the pulsar radio emission have fixed
amplitudes.

Pulsar radiation is modeled by a set of the noise circular
frequencies $\omega$ with the Gauss shape of the envelope. In the
model we specify a variation of a position angle (PA) $\chi$ along
the average profile of the pulse envelope. Influence of the
propagation medium is taken into account by eikonal equation
$\nabla \varphi(\omega) =n(\omega) \vec {k} (\omega)$, where
$\varphi (\omega)$ is the phase of the signal at circular
frequency $\omega$ , $n (\omega)$ is the ISM refractive
coefficient, $\vec {k} (\omega)$ is the wave vector. In this case
the equation for the phase of the analytical signal at arbitrary
frequency $\omega$ will have the form :
\begin{equation}
{\varphi (\omega)}_{O,X} \approx \omega \frac{L}{c} -
\frac{1}{\omega} \frac{2 \pi e^2}{m_e c} \int_0^L N_e(z)d z \mp
\frac{1}{\omega^2} \frac{2\pi e^3 }{{m_e}^2 c} \int_0^L N_e(z)
\beta(z) d z ,
\end{equation}
where $c$ is the speed of light, $e$ is the electron charge, $m_e$
is the electron rest mass, $L$ is the layer thickness, $N_e(z)$ is
the electron concentration on the line of sight, $\beta(z)$ is the
value of the projection of the magnetic induction vector onto that
line of sight.

The rotation of the polarization plane is connected with the different
refraction coefficients for the orthogonal modes that have the
opposite directions of rotation.
 These waves are the so-called ordinary ($O$) and extraordinary ($X$) waves
(\cite[Zheleznyakov 1997]{Zheleznyakov97}). 
For $O$ and $X$ waves refractive coefficients can be written as: $ n_{O,X}~=~ \sqrt {1-\left[{\omega _p}^2/\omega(\omega \mp \omega_H)\right]}$,
where $\omega_p$ and $\omega_H$ are the plasma and the cyclotron
frequencies, respectively, subscripts $O$ and $X$ correspond to
the ordinary ($-\omega_H$) and extraordinary ($+\omega_H$) waves.

An example of registering the polarization ellipse of PSR B0809+74
pulse on the radio telescope UTR-2 can be seen in the
Fig.\,\ref{fig1}a. The central recording frequency is $F_c = $
23.7 MHz. Period of the Faraday intensity modulation is $\Delta
F_F \approx$ 20 kHz. We have made the simulations for the similar
spectra. These spectra were generated for a given value of the
rotation measure ($RM =$ -234 rad/m$^2$) at frequencies $F_c$ = 20
MHz  and $F_c$ = 30 MHz  ($\Delta F =$  4 kHz). For the given
model signals the Faraday periods of intensity modulation were
found. Using these values the magnitude of the rotation measure
was estimated. The accuracy of this method is lower than
$0.5/N_{max}(f)$, where $N_{max}(f)$ is the number of harmonic,
which has the maximal intensity. This number depends on $\Delta
F_F$.

\section{Results}

We can construct a polarization tensor (\cite[Zheleznyakov
1997]{Zheleznyakov97}) for a signal registered in two orthogonal
polarizations. From polarization tensor the polarization
parameters (Stokes parameters) and the variation of PA can be
found (see Fig.\,\ref{fig1}).

In Fig.\,\ref{fig1}b, the Stokes parameters of the restored
signal (the continuous line) and the original signal (dashed line)
are presented. The variation of the PA in the pulse window at the
receiving end (Fig.\,\ref{fig1}c) is almost identical to the
variation of the model PA. The PA is determined only up to an
arbitrary constant.

The developed algorithms were used for processing the real data.
For three different anomalously intensive pulses of PSR B0809+74
registered near the center frequency $F_c$ = 23.7 MHz in the same
observation session of one hour duration the variations of the
values of the rotation measure of about 1 rad/m$^2$ were detected.
The values of the observed variations are 13 times larger than the
errors of determination of the rotation measure values. These
variations were observed during nighttime for undisturbed
ionosphere. Comparison of the $RM$ value obtained in the decameter
range with $RM$ values obtained in the 250-500 MHz frequency range
(\cite[Manchester 1972]{Manchester72}) allows us to suppose that
regular component of the interstellar medium magnetic field and
dipole component of the emitting pulsar pole of PSR B0809+74 have
opposite directions.

\section{Conclusions}

The direct and inverse problems of determining the polarization
parameters of elliptically polarized signal  propagating through
ISM were solved by numerical methods. An eikonal equation gives an
opportunity to take into account both the Faraday effect and cold
plasma influence on the signal properties.

A method of estimating the rotation measure and position angle was
proposed. It allows us to recover all Stokes parameters in the
reference frame associated with a pulsar.

It was found that ISM magnetic field and visible pulsar magnetic
field have opposite directions.

\end{document}